# A species of *Coprococcus* is related to BMI in patients who underwent malabsorptive bariatric surgery and its abundance is modified by magnesium and thiamin intake.

Fernando Suárez-Sánchez[a,*], Evelyn Pérez-Ruiz[a,b], Claudia Ivonne Ramírez-Silva[c], Mario Antonio Molina-Ayala[d], Sandra Rivera-Gutiérrez[e], Lizbel León-Solís[e], Lázaro García-Morales[f], Arturo Rodríguez-González[d], César Martínez-Ortiz[d], Luis Axiel Meneses-Tapia[a,] Miguel Cruz-López[a].

[a] *Medical and Biochemistry Research Unit, Specialty Hospital, Centro Médico Nacional Siglo XXI, Instituto Mexicano del Seguro Social, Mexico City, Mexico.*

[b] *Faculty of Chemistry, Doctorate Program in Medical, Dental and Health Sciences, Clinical and Experimental Health Research, Universidad Nacional Autonoma de México, Mexico City, Mexico.*

[c] *Department of Maternal, Children and Adolescent Nutrition. Nutrition and Health Research Center, Instituto Nacional de Salud Pública, Morelos, Mexico.*

[d] *Diabetes and Obesity Clinic, Specialty Hospital, Centro Médico Nacional Siglo XXI, Instituto Mexicano del Seguro Social, Mexico City, Mexico.*

[e] *Department of Microbiology, Escuela Nacional de Ciencias Biológicas, Instituto Politécnico Nacional, Mexico City, Mexico.*

[f] *Department of Molecular Biomedicine, Centro de Investigación y de Estudios Avanzados del Instituto Politécnico Naciona, Mexico City, Mexico.*

This work was supported by the National Council of Science and Technology (CONACYT-México) with the grant SALUD-2017-C02-289961

[*] Corresponding author: Fernando Suárez Sánchez, Medical and Biochemistry Research Unit, Specialty Hospital, Centro Médico Nacional Siglo XXI, Instituto Mexicano del Seguro Social, Mexico City, Mexico, tel. 55 5627 6900, fs.bioq.imss@gmail.com

Authors' e-mail addresses: Fernando Suárez-Sáncheza (fs.bioq.imss@gmail.com), Evelyn Pérez-Ruiz (evepuma13@gmail.com), Claudia Ivonne Ramírez-Silva (ciramir@insp.mx), Mario Antonio Molina-Ayala (mmol_17@yahoo.com.mx), Sandra Rivera-Gutiérrez (san_rg@yahoo.com.mx), Lizbel León-Solís (lizbelleon@gmail.com), Lázaro García-Morales (laz_gm@hotmail.com), Arturo Rodríguez-González (arturorodriguezmd@gmail.com), César Martínez-Ortiz (camartinez@me.com), Luis Axiel Meneses-Tapia (qcmeneses90@hotmail.com), Miguel Cruz-López (mcruzl@yahoo.com).

**Abstract**

*Background*: Morbid obesity is associated with metabolic alterations and the onset of type 2 diabetes. Patients who undergo a malabsorptive bariatric surgery show an important improvement in several clinical variables and a modification in the gut microbiota balance. In this study, we aimed to identify bacteria related to changes in the body mass index of patients who underwent a bariatric surgery and their relationship with nutrients intake.

*Results*: There were differences in bacterial diversity in the gut microbiota of patients that underwent a bariatric surgery. The Shannon and Simpson indexes decrease after the surgery ($p < 0.001$) and the beta diversity indexes (Bray-Curtis, Weighted and Unweighted UniFrac) showed differences when comparing pre- and post-surgery ($p = 0.001$). The abundance of a species in the genus *Coprococcus* correlated positively with the intake of magnesium and thiamin in post-surgery individuals (rho = 0.816, $p_{FDR}$ = 0.029 and rho = 0.812, $p_{FDR}$ = 0.029, respectively) and was related to BMI in both groups ($p = 0.043$ pre-surgery and $p = 0.036$ post-surgery). The abundances of several bacteria belonging to the order *Clostridiales,* as well as an enrichment of vitamin B1 (thiamin) biosynthesis, sugar degradation, acetate production and some amino acids biosynthesis were higher before the surgery.

*Conclusions*: The abundance of a species of the genus *Coprococcus* that showed inverse relationships with BMI in pre-surgery and post-surgery patients correlates with the intake of magnesium and thiamin in individuals that underwent a malabsorptive bariatric surgery. It indicates that the well-established beneficial effects of bariatric surgery on BMI may be amplified by modulating the intake of micronutrients and its effect on the gut bacterial.

# 1. Introduction

Obesity is a disease characterized by the excessive accumulation of fat at several body sites. Deposits of fat in the waist are the most detrimental and are associated with the onset of other pathologies such as cardiovascular disease and type 2 diabetes (T2D) [1]. A body mass index (BMI) equal to or greater than 40 is classified as morbid obesity [2] . In addition to the inherent metabolic alterations that accompany obese patients, profound effects on their mobility and everyday life appear. In most cases, dietary and exercise interventions have limited impact on weight loss; thus, other approaches, such as bariatric surgery, are implemented [3]. Although different types of bariatric surgery are available, the use of malabsorptive bariatric surgery in obese patients has shown very positive outcomes. The most common type of bariatric surgery is Roux-en-Y gastric bypass (RYGB) but one-anastomosis gastric bypass (OAGB) has also been used successfully. In both cases, intestinal rearrangement limits the nutrients absorbed by bypassing the proximal portion of the small intestine and facilitating the passage of poorly digested food to the distal small intestine and colon [4]. Those changes contribute to modifications of the bacterial diversity and balance in the gut. This new balance has an effect on the capacity of gut bacteria to perform biologically relevant functions such as the synthesis of essential vitamins, the production of molecules with signaling effects and the digestion of nutrients [5].

Prebiotics and probiotics have been used to modulate the abundance of some bacteria that are considered beneficial for the host. For example, fiber intake has been extensively proposed to increase the growth of short-chain fatty acid (SCFA)-producing bacteria[6]. However, there is still a need to understand how other dietary micronutrients can modulate the gut flora in such a way that clinical variables improve. In this study, we identified a species of *Coprococcus* whose abundance is greater in patients with larger BMI after the surgery. The response of this bacteria to dietary intake was subsequently investigated, and we identified two micronutrients (magnesium and thiamin) that correlated with its abundance in the gut. Changes in several other bacteria and bacterial metabolic pathways were observed in patients who underwent malabsorptive bariatric surgery.

## 2. Materials and Methods

### 2.1. Study population

Participants were recruited at the Specialty Hospital, National Medical Center Century XXI, Mexican Social Security Institute. All participants (21 patients) were classified as morbidly obese and were programmed to receive elective malabsorptive bariatric surgery (Roux-en-Y gastric bypass (RYGB) or one-anastomosis gastric bypass (OAGB)). Patients were excluded if they had undergone a previous cholecystectomy or bariatric surgery, if they had taken antibiotics in the previous three months or if they showed any sign of gastrointestinal infection. The patients signed an informed consent form prior to the collection of any blood or fecal samples, anthropometric measurements, or interviews about their dietary intake. This study was approved by the National Committee of Ethics, Mexican Social Security Institute, ensuring that it conformed to the ethical guidelines of the Declaration of Helsinki.

### 2.2. Sample collection and diet

Blood samples were drawn before and 109 ± 28 days after bariatric surgery. Patients were instructed to fast for 10 hrs and serum was used to measure the concentrations of glucose, triglycerides, total cholesterol, HDL, LDL and insulin. Fecal samples were received, aliquoted and stored at -70°C by laboratory personnel on the same days that the blood samples were drawn. A 7-day dietary recall questionnaire validated and used in the National Survey of Health and Nutrition (ENSANUT) and complemented with seasonal fruit and vegetables was used during the interrogation to estimate nutrient intake [7, 8]. The questionnaire collected information on the intake of 191 food items divided into 16 sections.

*2.3. Micro- and macronutrient estimations*

Energy and nutrient intake were calculated according to the methodology used in the ENSANUT [9] with the nutrient database constructed with items from the Mexican Food Database (MFD). This database is a compilation of several food composition databases prepared by the Center for Nutrition and Health Research of the National Institute of Public Health [10]. The survey was performed by nutritionists and the data were transformed to micro- and macronutrient intake per day. Data processing was performed in Stata v.14.0 (Stata Corporation).

*2.4 DNA extraction and 16S rDNA sequencing*

Fecal samples were collected by the patients no more than 24 h prior their delivery in the laboratory, after which the samples were aliquoted and stored at -80°C until use. DNA was extracted from 150 to 200 mg of stools using the QIAamp DNA Mini Kit (Qiagen) according to the manufacturer's instructions, except that a step was added at the beginning of the extraction procedure to mechanically rupture bacteria in the samples with the TissueLyser LT equipment (Qiagen) for 5 minutes at 60 cycles/second. DNA integrity was evaluated by electrophoresis in 2% agarose gels and purity was estimated by measuring the 260/280 ratio in a BioTek Epoch™ Microplate Spectrophotometer (Agilent Technologies).

Library preparation and sequencing of the hypervariable region V4 of the bacterial 16S rDNA gene were performed based on the methodology described by Kozich Westcott [11]. Briefly, 20 ng of DNA was quantified in a Qubit™ (Thermo Fisher Scientific) using the Quant IT™ dsDNA HS Assay Kit. This was used as a template for the amplification (25 amplification cycles) of the V4 region using the AccuPrime™ Pfx SuperMix high-fidelity DNA polymerase from Invitrogen™ (Thermo Fisher Scientific). The product was purified with Beckman Coulter™ AMPure XP 1.8X beads (Life Sciences) and the amplicon size was assessed with the 4200 TapeStation High Sensitivity DNA Reagent Kit (Agilent Technologies). The DNA concentration was measured with a Qubit™ instrument and the Quant IT™ dsDNA HS Assay Kit (Thermo Fisher Scientific) and the concentration was adjusted to 4 nM. Pooling of the samples was followed by a new quantification using the Qubit™ and the Quant IT™ dsDNA HS assay kit (Thermo Fisher Scientific).

The equimolar DNA pool was denatured with NaOH and spiked with PhiX (PhiX Control Kit v3 Illumina). Paired-end DNA sequencing was performed on a MiSeq system (Illumina) with a MiSeq Reagent Kit V3 (600-cycle) (Illumina). All the oligonucleotide sequences used were obtained from Kozich, Westcott [11].

## 2.5 Data analysis

Sequencing quality control and identification of the amplicon sequence variants (ASVs) were performed in QIIME2 and the DADA2 package [12, 13]. QIIME2 was used to calculate alpha and beta diversity and taxonomic assignment was achieved by using the Greengenes database [14]. Comparisons of phenotypic variables and alpha diversity between groups were performed using Student's t test or the Mann–Whitney U test on normally distributed or nonnormally distributed data respectively. Statistical distribution of the variables was assessed with the Shapiro–Wilk test. PERMANOVA was employed to assess differences in Bray–Curtis, weighted and unweighted UniFrac data. Spearman correlation analysis with p adjusted by false discovery rate (FDR) was used to identify relevant correlations between bacterial abundance, energy, and nutrient intake (for macro- and micronutrients). The abundance of one species of the genera *Coprococcus* (named g_*Coprococcus*;s_ from hereon) in the pre- and postsurgery groups was stratified as "higher than" or "lower than" the median abundance per group. The BMI was then compared between the newly created dichotomized variables in the pre- and postsurgery groups. Bacterial abundance differences between groups were calculated using the Mann–Whitney U test adjusted by FDR. Picrust2 analysis was applied to the species-level relative abundance tables to identify enriched metabolic pathways [15]. The function aldex in the ALDEx2 library in R (version 4.3.0 [2023-04-21 ucrt]) was used to compute the effect size. R was employed to perform statistical analysis and plotting as well.

## 3. Results

### *3.1 Phenotypic characterization*

We observed an improvement in anthropometric and biochemical parameters after the malabsorptive bariatric surgery was performed in patients with obesity. In particular, the patients lost an average of 28 kg, which represents approximately 24% of their initial weight. The majority of patients were reclassified into the obesity class I group ($30 \leq BMI < 35$) instead of the morbid obesity group ($BMI \geq 40$) where they were initially allocated. Serum levels of glucose, triglycerides, LDL and total cholesterol decreased after surgery (Table 1).

Table 1: Anthropometric and biochemical characteristics of the population.

|  | **General<br>n = 42** | **Presurgery<br>n = 21** | **Postsurgery<br>n = 21** | **p value** |
|---|---|---|---|---|
| Age (years) | 44.7 ± 7.0 | 44.7 ± 7.1 | 44.7 ± 7.1 | 1 |
| Gender (female) | 26 (62%) | 13 (62%) | 13 (62%) | 1 |
| Weight (kg)* | 106.1 [86.2 - 130.4] | 117.0 [102.0 - 139.5] | 88.9 [77.8 - 116.7] | **0.0061** |
| BMI* | 38.8 [33.7 - 46.6] | 42.6 [39.4 - 47.2] | 33.7 [30.1 - 37.0] | **0.0004** |
| Waist (cm) | 126.4 ± 24.6 | 132.5 ± 24.0 | 119.6 ± 24.3 | 0.1206 |
| Hip (cm) | 134.4 ± 20.5 | 139.0 ± 19.9 | 129.2 ± 20.5 | 0.1575 |
| WHR | 0.94 ± 0.11 | 0.95 ± 0.09 | 0.93 ± 0.13 | 0.5128 |
| Glucose (mg/dL)* | 84.4 [79.3 - 90.8] | 87.8 [84.7 - 93.7] | 80.0 [78.9 - 84.3] | **0.0041** |
| Triglycerides (mg/dL)* | 103.4 [88.1 - 146.9] | 134.6 [96.1 - 168.2] | 95.8 [87.6 - 118.3] | **0.0382** |
| HDL (mg/dL) | 43.4 ± 12.4 | 46.1 ± 10.1 | 40.6 ± 14.1 | 0.1561 |
| LDL (mg/dL)* | 102.0 [77.3 - 122.0] | 112.3 [97.6 - 143.0] | 88.6 [67.3 - 103.5] | **0.0044** |
| Cholesterol (mg/dL) | 166.6 ± 46.8 | 187.2 ± 42.3 | 146.0 ± 42.5 | **0.0031** |

* Mann–Whitney U test was applied when the variable showed a nonnormal statistical distribution. Otherwise, Student's t test was used. Fisher's test was applied to compare sex distributions. WHR: Waist-to-hip ratio.

*3.2 Alpha and beta diversity*

Microbiota richness assessed by the Chao1 index was not significantly different before or after surgery (fig 1a). However, the Shannon and Simpson indices differed between the two groups (p values of $5 \times 10^{-4}$ and $8 \times 10^{-4}$, respectively), with a greater diversity index observed in the presurgery group (Figure 1c and e). On the other hand, beta diversity was different for the Bray–Curtis, weighted and unweighted UniFrac indices. In all three cases, the PERMANOVA p value was 0.001 (fig 1b, d, f).

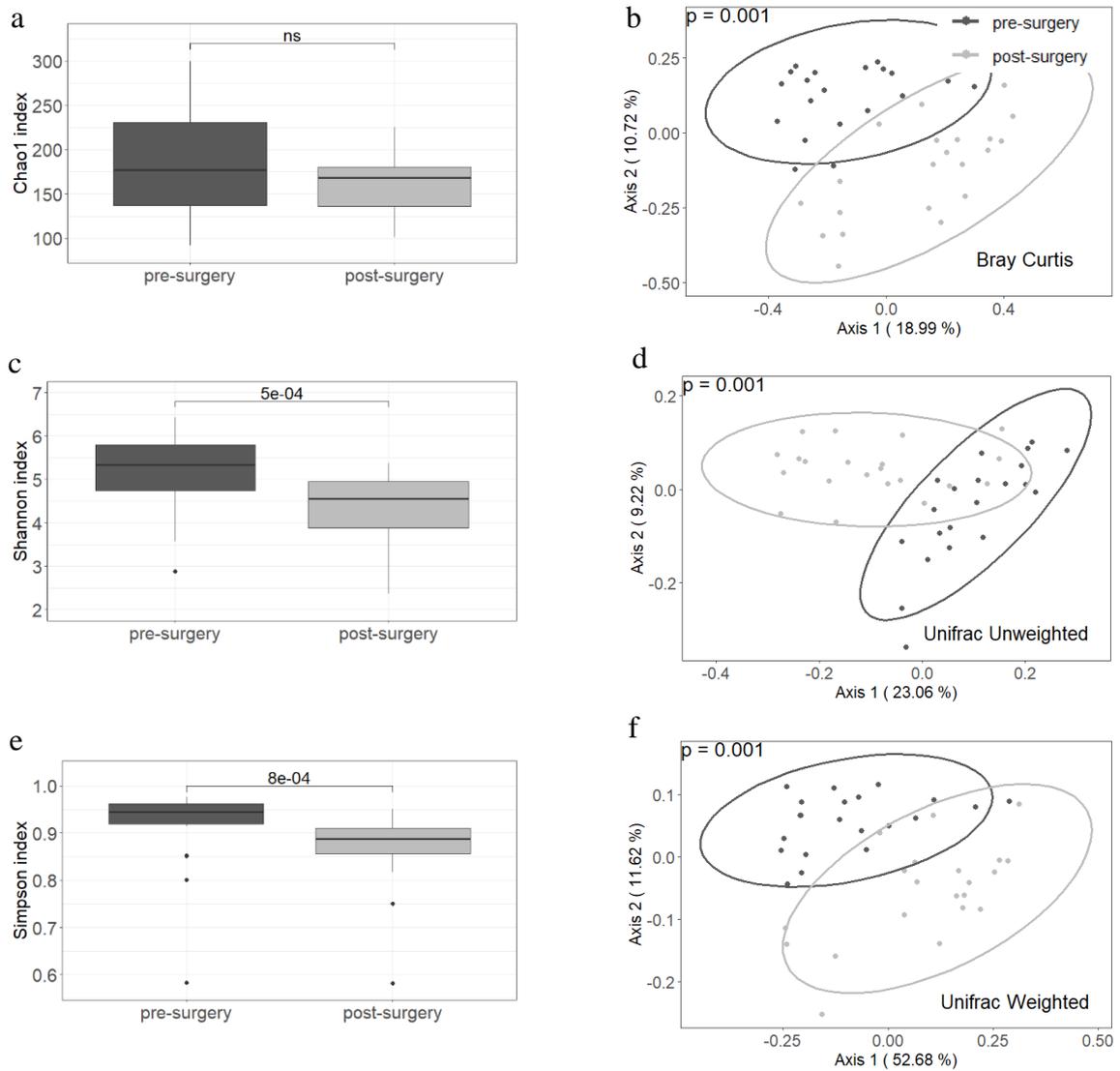

Figure 1: Alpha and beta diversity are different before and after a malabsorptive bariatric surgery. The Chao1 index (a) was not statistically different between groups but Shannon and Simpson indexes (c, e) decreased after the bariatric surgery. Bray-Curtis, Unweighted and Weighted Unifrac are all different when compared before and after the bariatric surgery (b, d, f).

*3.3    Correlation between the microbiota and nutrient intake*

Carbohydrate, protein, and total energy intake showed a trend to decrease in the postsurgery group; however, the difference did not reach statistical significance. In contrast, the intakes of magnesium and thiamin (vitamin B1) were positively correlated with the relative abundance of g_*Coprococcus*;s_ in the postsurgery group (rho = 0.82, $p_{FDR}$ = 0.029; rho = 0.81, $p_{FDR}$ = 0.029, respectively) (Figure 2). Further comparison of anthropometric, biochemical and energy intake variables between the low (lower or equal to the median) and high (higher than the median) g_*Coprococcus*;s_ relative abundance groups showed that BMI was greater in patients with higher g_*Coprococcus*;s_ relative abundance (p = 0.036; Fig. 2c). This relationship was reversed in the presurgery group of patients in which the BMI decreased while the relative abundance of g_*Coprococcus*;s_ increased (p = 0.043, Fig 2d). When the energy intake was compared by g_*Coprococcus*;s_ abundance, it was greater (2066.4 kcal [1497.9 kcal – 2262.8 kcal]) in the higher abundance group than in the low abundance group (1379.6 kcal [1007.8 kcal – 1680.1 kcal], p = 0.024) only in bariatric surgery patients, but this difference was absent in the presurgery group of patients.

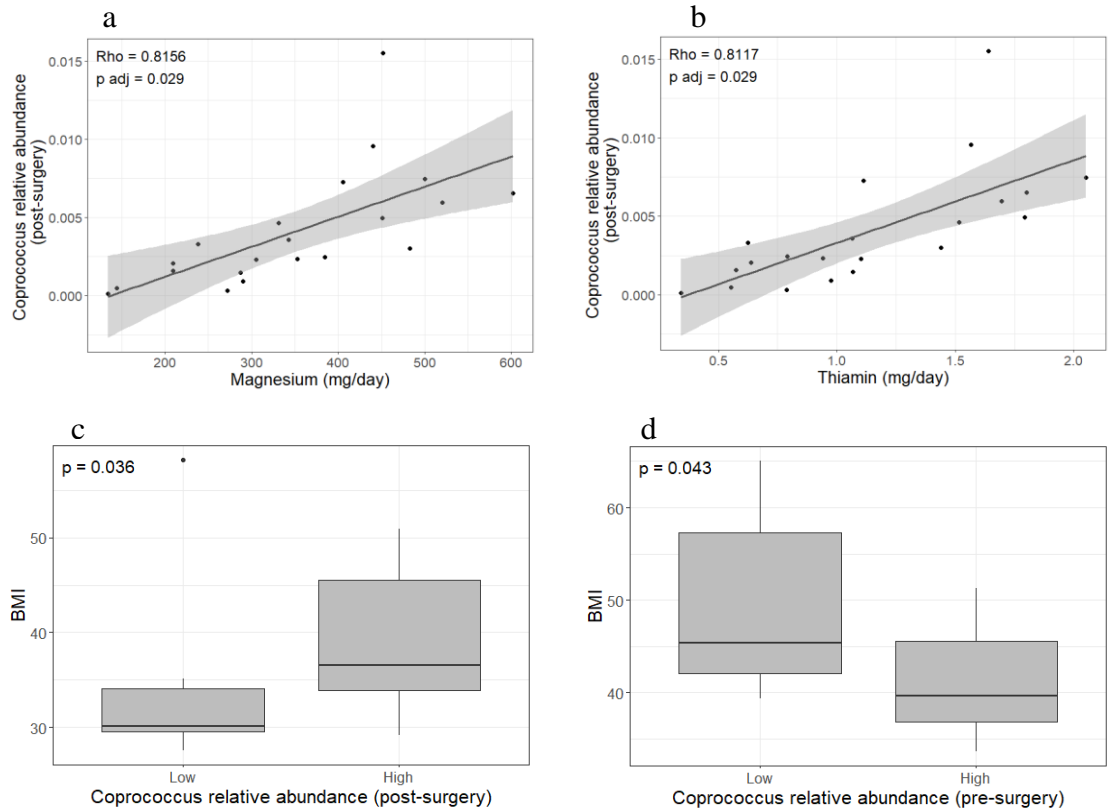

Figure 2: The *Coprococcus* abundance correlates with micronutrients intake after patients undergo a malabsorptive bariatric surgery and inversely relates to the BMI before and after the surgery. Intake of magnesium and thiamin positively correlates with *Coprococcus* abundance in the post-surgery group (a, b) but not in the pre-surgery group (data not showed). Patients with low *Coprococcus* abundance (dichotomized in lower and higher quantiles) are characterized by a lower BMI in the post-surgery group (c) but the opposite was observed in the pre-surgery group (d).

*3.4 Changes in the microbiota before and after the bariatric surgery*

Eighteen bacterial species (which complied with a $p_{FDR} \leq 0.05$ and a median count of 50 sequences in the rarefied abundance data across the studied population) exhibited different relative abundances between the groups before and after bariatric surgery (Figure 3). Most bacteria decreased in abundance after surgery (all belonging to the order *Clostridiales*), but the abundance of *Streptococcus*, *Megasphaera*, *Veillonella dispar* and *Veillonella parvula* increased postsurgery. Specifically, bacteria that differed between groups belonged to the order *Clostridiales* and were grouped into three families: *Lachnospiraceae*, *Ruminococcaceae* and *Veillonellaceae*. At the phylum level, *Bacteroidetes* and *Fusobacteria* increased after surgery (0.32 [0.17-0.43] vs. 0.52 [0.42-0.60], $p_{FDR} = 0.014$ and 0.0 [0.0-0.0] vs. 0.0 [0.0-0.02], $p_{FDR} = 0.002$; respectively), while *Firmicutes* decreased (0.61 [0.49-0.73] vs. 0.38 [0.30-0.48], $p_{FDR} = 0.002$) (supplementary data S1).

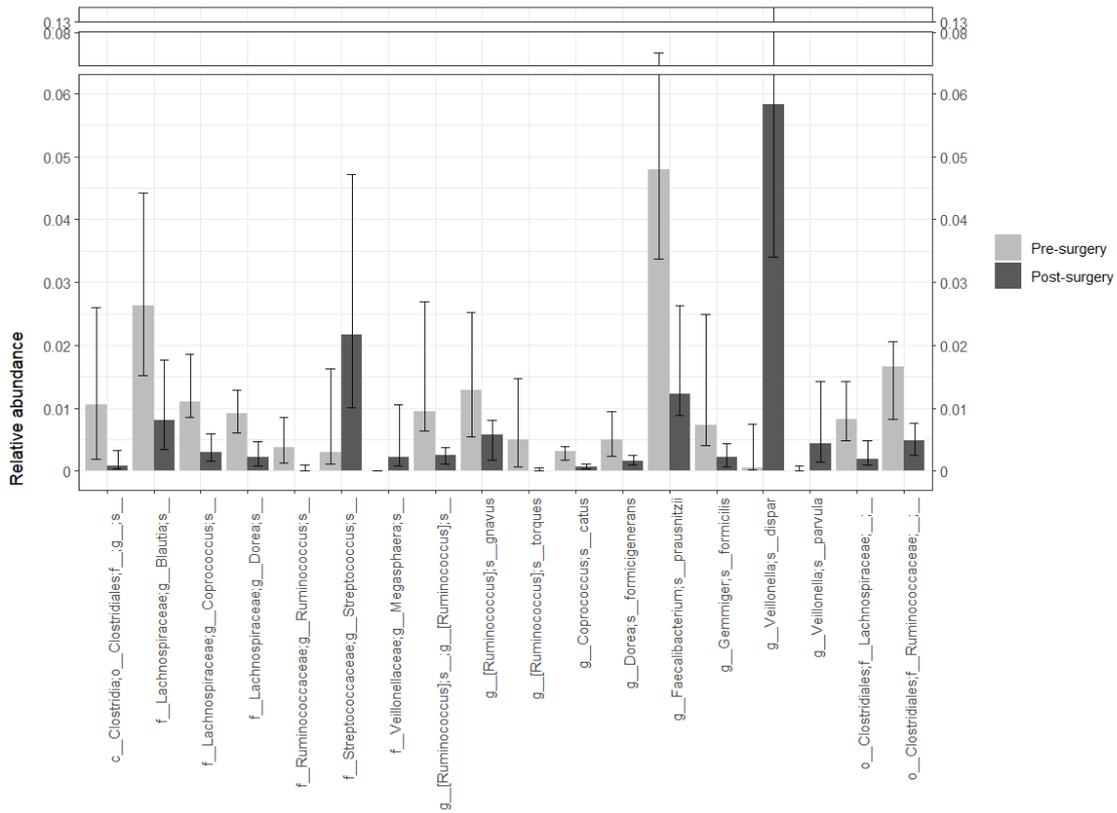

Figure 3: Bacterial abundance changes after patients undergo a malabsorptive bariatric surgery. An abundance reduction in most bacteria was observed (pFDR < 0.05). Genera *Streptococcus*, *Megasphaera* and species *Veillonella dispar* and *Veillonella parvula* are more abundant in the post-surgery group. Only bacteria that had a median abundance count higher than 50 across the studied population is shown.

*3.5    Enriched bacterial metabolic pathways*

The enrichment of bacterial metabolic pathways is presented in Figure 4. The main pathways enriched in the postsurgery group were related to the synthesis of vitamin K. On the other hand, in the presurgery group, most identified enriched pathways were involved in amino acid (lysine, ornithine, arginine, histidine, serine, glycine and isoleucine) and vitamin B1 (thiamin) biosynthesis, sugar (sucrose, galactose, fucose, rhamnose, glucose, xylose and mannan) and starch degradation, acetate production and glycogen biosynthesis and degradation.

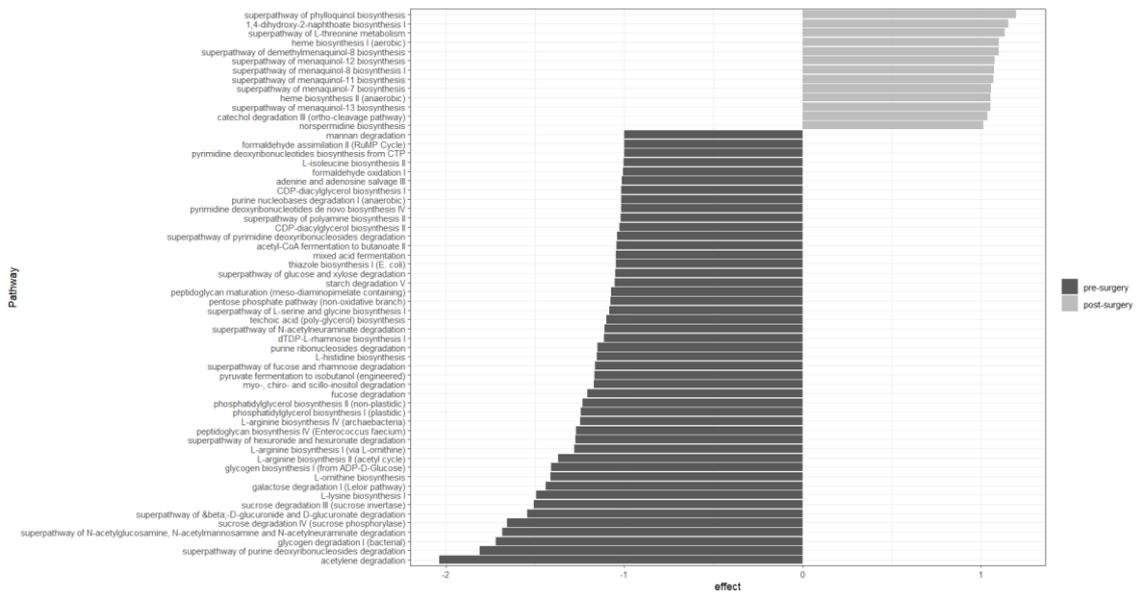

Figure 4: Bacterial metabolic pathways enriched in the pre and post-surgery groups are related to amino acid, vitamin B1 (thiamin) and K biosynthesis, sugar degradation and acetate production.

# 4. Discussion

Bariatric surgery is a procedure that has been shown to importantly reduce weight and improve metabolic alterations associated with diabetes. Major remodeling of the digestive tract contributes to these changes by increasing the release of anorexogenic hormones such as PYY and GLP-1, decreasing nutrient absorption but also altering the gut microbiota composition (Madsbad et al., 2014). This study revealed that bacterial diversity (alpha and beta) varies greatly with several species decreasing in abundance after bariatric surgery. The abundance of one of those bacteria (g_*Coprococcus*;s_) showed an opposite relationship with BMI before and after surgery and was positively correlated with the intake of magnesium and thiamin only in the postsurgery group. These results indicate that after surgery, g_*Coprococcus*;s_ is positively related to BMI and is responsive to nutrient intake. It opens the possibility to future exploration of the feasibility of the modulation of the g_*Coprococcus*;s_ abundance by modifying micronutrients intake in order to amplify the weight loss and to augment the metabolic benefits that accompany the surgery.

Overall, the results of this study are in agreement with previously published findings showing that bariatric surgery improves patient health. Patients in the present study lost an average of 24% of their initial weight and they were reclassified as obese class I instead of their initial classification as morbidly obese. This represents an important reduction in weight. In addition, at the beginning of the study, 71.4% of the patients were taking some type of hypoglycemic medication, but this figure was reduced to 14.3% after surgery. Other improvements observed were in biochemical variables such as fasting glucose, triglycerides, LDL and total cholesterol (Table 1), which reduce the risk of worsening metabolic alterations as well as the onset of cardiovascular events that have been widely associated with obesity and T2D.

The improvement in anthropometric and biochemical parameters was accompanied by a decrease in the Shannon and Simpson diversity indices (Figure 1). This reduction in diversity is attributed to the extensive remodeling of the gut, which prevents the transit of food through the first portion of the small intestine as well as incomplete digestion. The beta diversity estimated by the Bray–Curtis, unweighted and weighted UniFrac indices showed a clear difference between the compared groups (Figure 1). These results indicate that the bacterial composition is strongly modified after the malabsorptive bariatric surgery. We further analyzed the data to identify bacteria whose abundance changed after surgery. At the phylum level, the abundance of *Firmicutes* decreased, while that of *Bacteroidetes* and *Fusobacteria* increased (supplementary material S1). These findings are in agreement with reports in the literature indicating that the microbiota of obese patients is enriched in *Firmicutes* (Indiani et al., 2018). At lower taxonomical levels and after adjusting the p value by false discovery rate, we identified 18 taxa that varied in abundance (Figure 3). Most taxa decreased in abundance after bariatric surgery (14 out of 18) which is in agreement with the lower alpha diversity in the postsurgery group and may account for part of the differences observed in beta diversity. These taxa belong to the order *Clostridiales* and the families *Lachnospiraceae* and *Ruminococcaceae*. The genomes of members of both families are known to contain both ATP-binding cassette (ABC) genes and other genes that produce enzymes that participate in the degradation of starch and sugars such as xylose, galactose, melibiose, lactose, cellulose, L-arabinan, N-acetylmannosamine and N-acetylneuraminate (Biddle et al., 2013). The hydrolysis of these sugars results in the production of butyrate, acetate, and propionate, which are SCFAs that participate in gut homeostasis and energy harvesting (Biddle et al., 2013). In this study, we identified enrichment of several pathways involved in starch and sugar degradation (i.e. xylose, galactose, N-acetylmannosamine, N-acetylneuraminate, N-acetylglucosamine, mannan, glucose, rhamnose, fucose, sucrose) in obese patients before bariatric surgery (Figure 4). Although an increase in the production of SCFAs is considered a beneficial trait that promotes gut homeostasis, if combined with a greater capacity of bacteria to extract energy from the diet, this might result in an increase in the obese phenotype.

Among the pathways that showed differences between groups, the acetylene degradation pathway was the most enriched in the presurgery group. The end product of this pathway is microbiota-derived acetate. Although deficiency of this SCFA has been associated with learning and memory impairment (Zheng et al., 2021), oxidative stress (Olaniyi et al., 2022), inflammation (Mandaliya et al., 2021)(Daïen et al., 2021) and other metabolic alterations (Canet et al., 2022), (Olaniyi et al., 2021), (Olaniyi & Amusa, 2020), there is evidence indicating that microbiota acetate production due to high fructose consumption (which is relatively common in obese patients) leads to the activation of the hepatic ACSS2 enzyme which transforms acetate to lipogenic acetyl-CoA (Zhao et al., 2020). This occurs in addition to the more studied pathway in the liver that involves the conversion of glucose and fructose to fatty acids with the participation of citrate and the enzyme ACLY (Softic et al., 2017). Thus, the observed enrichment of the microbiota acetylene degradation pathway in obese patients before surgery combined with high caloric content diets might contribute to the maintenance of the obese phenotype.

Other energy-related pathways enriched in the presurgery group were the microbial glycogen degradation and glycogen biosynthesis pathways. This finding indicates that the microbiota of morbidly obese patients has the capacity to store energy surplus due to the patients' intake of high caloric diets.

A deficiency of vitamin B1 (thiamin) has been previously reported in approximately 27% of patients who underwent bariatric surgery (Bahardoust et al., 2022). This is significant since those patients have to receive vitamin B1 supplements for the rest of their lives. This essential vitamin is obtained from the diet and as a product of bacterial metabolism. Vitamin B1 is crucial and indirectly involved in the electron transport chain and extraction of energy from carbohydrates (Manzetti et al., 2014). Low levels of this vitamin cause disorder such as beriberi syndrome, Wernicke–Korsakoff syndrome, optic neuropathy and Leigh's disease and include symptoms such as malaise, weight loss, irritability and confusion (Smith et al., 2021). We found that the abundance of bacteria involved in the thiazole biosynthesis pathway was lower in postsurgical patients (Figure 4). The combination of these findings with lower absorption rates may explain the deficiency of this vitamin after bariatric surgery. Thiamin is absorbed primarily in the proximal part of the small intestine which is the region that is hindered during bariatric surgery (Nath et al., 2017).

The family *Ruminococcaceae* is known to require thiamine for survival, but they lack all the genes necessary for its novo synthesis (Park et al., 2022). Thus, the abundance of members of the *Ruminococcaceae* family is linked to the availability of this vitamin in the gut. The results presented in this study support these findings since we observed that the relative abundances of *Faecalibacterium prausnitzii*, *Gemmigers formicilis* and *Ruminococcus* (all of which belong to the *Ruminococcaceae* family) decreased after bariatric surgery (Figure 3). The analysis of the enrichment of bacterial metabolic pathways predicts a reduced abundance of genes required in the thiazole biosynthesis pathway in samples of patients after malabsorptive bariatric surgery (Figure 4). *Faecalibacterium prausnitzii* is considered to be a probiotic (Chang et al., 2019), (Dudík et al., 2022) due to its association with the levels of indoleproprionic acid (a potent antioxidant) (Menni et al., 2019), butyrate production, regulation of intestinal barrier permeability (Xu et al., 2020), (Moosavi et al., 2020) and a reduction in circulating BCAA levels and insulin sensitivity (Moran-Ramos et al., 2021). Thus, fine regulation of thiamin levels and fiber intake in postbariatric surgery patients is important for maintaining a healthy abundance of *Faecalibacterium prausnitzii* and reducing thiamine deficiency-associated disorders. However, since we identified that g_*Coprococcus*;s_ abundance is related to the BMI after surgery and is responsive to the intake of thiamine, the intake of thiamin must be such that it do not have a detrimental effect on BMI mediated by g_*Coprococcus*;s_ (Figure 2).

*Coprococcus catus* and another unidentified species of the *Coprococcus* genus are other SCFA-producing bacteria whose abundance decreased after bariatric surgery. Furthermore, the abundance of g_*Coprococcus*;s_showed an inverse relationship with BMI in patients before and after bariatric surgery (Figure 2). Before the surgery, obese patients with higher gut content of this bacteria had a lower BMI. Previous studies have shown that *Coprococcus* is associated with improved insulin sensitivity in metabolically healthy obese individuals (Cui et al., 2022), (Zhang et al., 2022), (Olivares et al., 2021). In addition, *Coprococcus* species have been shown to decrease inflammation, enhance anti-inflammatory molecules and mucin production and restore tight junctions (Huang et al., 2022), (Yang et al., 2023).

In contrast to these findings in patients before surgery, a higher abundance of g_*Coprococcus*;s_ was related to a higher BMI after surgery. This was surprising and opposite to the abovementioned reports of the relationship between g_*Coprococcus*;s_ and BMI before the surgery. A study by Lozano et al. (Lozano et al., 2022) reported that a positive association between high inflammatory index diets and ectopic fat accumulation was observed only in patients with lower gut *Coprococcus* abundance.

The intake of high-inflammatory-index diets with high caloric content is frequent in obese patients but less common in postsurgical patients which may explain why a relationship between lower g_*Coprococcus*;s_ abundance and BMI was observed in presurgical subjects but not in patients who had already undergone surgery. In addition, we observed that postsurgery patients in the lower abundance quantile reported a lower caloric consumption than patients in the higher abundance quantile (1379.6 kcal [1007.8 kcal -1680.1 kcal] vs. 2066.4 kcal [1497.9 kcal -2262.8 kcal], p = 0.024). We further observed that the relative abundance of g_*Coprococcus*;s_ in the postsurgery group was positively correlated with magnesium and thiamin intake. Both micronutrients are crucial in the oxidative metabolism of glucose (LeBlanc et al., 2017). Indeed, magnesium is required for the proper activation of vitamin B1 to thiamine diphosphate (TDP) (Piuri et al., 2021). It would be relevant to further explore the relationships between BMI, g_*Coprococcus*;s_abundance, magnesium and thiamin intake and diet supplementation to determine the reasons underlying the divergent relationship between BMI and g_*Coprococcus*;s_abundance as a consequence of bariatric surgery.

The abundance of only four bacteria belonging to the families *Streptococcaceae* and *Veillonellaceae* (*Streptococcus*, *Megasphaera*, *Veillonella dispar* and *Veillonella parvula*) increased after surgery. In particular, the genus *Veillonella* is a common oral bacterium that is positively associated with caries and fried meat consumption (Rosier et al., 2022), (Gao et al., 2021). Its enrichment in the gut of postsurgical patients may be related to a more direct passage of food to the intestines and reduced digestion by acids in the stomach. It has been reported that bacteria of the genus *Veillonella* can produce vitamin K (Ramotar et al., 1984). Thus, the enrichment of several menaquinol biosynthesis pathways after surgery might be the result of a greater abundance of *Veillonella* in the gut.

Other metabolic pathways enriched before surgery were involved in inositol degradation and amino acid biosynthesis (lysine, ornithine, arginine, histidine, serine, glycine and isoleucine). Three of those amino acids are considered essential (lysine, histidine, isoleucine).

## 5. Conclusion

Important changes in clinical variables and the microbiota are observed after individuals with morbid obesity undergo malabsorptive bariatric surgery. Only 20% of patients who took hypoglycemic medication before surgery continued doing so after surgery, and the remaining 80% did not require medication to control their glucose levels. This improvement was accompanied by changes in the gut microbiota alpha and beta diversity. The microbiota after surgery was less diverse, with a decrease in the abundance of several bacteria. These changes in the microbiota profile drove a reduction in the enrichment of metabolic pathways involved in amino acid, vitamin B1 (thiamin), sugar degradation and acetate production. In particular, a decrease in the enrichment of sugar degradation pathways after surgery may contribute to a reduction in the harvesting and storing energy capacity enhanced in obese patients. A relationship between BMI and the abundance of g_*Coprococcus*;s_ as well as a positive correlation between the dietary intake of magnesium and thiamine and the same bacteria after surgery was identified.

## Declarations

*Ethics approval and consent to participate*

This study was approved by the National Committee of Ethics, Mexican Social Security Institute, with authorization number R-2017-785-063. Participants signed an informed consent before any sample was obtained or clinical data recorded.

*Consent for publication*

N/A. Manuscript does not contain individual person's data.

*Data availability*

All data generated or analyzed during the study are included in this published article (and its supplementary information files). The sequencing raw data was uploaded at the NCBI repository and the Bioproject and associated SRA metadata can be accessed at [https://dataview.ncbi.nlm.nih.gov/object/PRJNA1181870?reviewer=vtmoq00cnbsha6812mss5gs4iu](https://dataview.ncbi.nlm.nih.gov/object/PRJNA1181870?reviewer=vtmoq00cnbsha6812mss5gs4iu). Any further request for the data used or generated in this study please contact Dr. Fernando Suarez-Sanchez at [fs.bioq.imss@gmail.com](mailto:fs.bioq.imss@gmail.com).

*Declaration of competing interests*

The authors declare that they have no known competing financial interests or personal relationships that could have appeared to influence the work reported in this paper.

*Funding*

This work was supported by the National Council of Science and Technology (CONACYT-México) with the grant SALUD-2017-C02-289961. CONACYT grants to Evelyn Pérez-Ruiz: 368994.

*Authors' contributions*

EPR: Formal Analysis, Writing – review & editing. CIRS: Formal Analysis. MAMA: Provided essential material. SRG: Methodology. LLS: Formal Analysis, Writing – original draft, Writing – review & editing. LGM: Methodology. ARG: Provided essential material. CMO: Provided essential material. LAMT: Methodology. MCL: Provided essential reagents. FSS: Conducted Research, Designed Research, Formal Analysis, Supervision, Writing – review & editing.

*Acknowledgments*

We thank Iliana Avila Soto and Sofia Barragan for their assistance during the application and nutrient calculations from the diet questionnaires as well as Carolina González and Javier Gaytán from the Sequecing Laborary, Centro de Instrumentos , División de Desarrollo de la Investigación, Instituto Mexicano del Seguro Social, for their technical assistance during the bacterial DNA sequencing.

# Appendix A. Supplementary data

The following is the Supplementary data to this article:

Supplementary data 1.

Table S1: Quartiles and adjusted p values of phyla identified in patients before (labeled as 1) and after (labeled as 2) the bariatric surgery.

| Phyla | P FDR | 1 - Q25% | 1 - Q50% | 1 - Q75% | 2 - Q25% | 2 - Q50% | 2 - Q75% |
|---|---|---|---|---|---|---|---|
| Actinobacteria | 0.204 | 0.005 | 0.006 | 0.020 | 0.003 | 0.004 | 0.009 |
| Bacteroidetes | **0.014** | 0.170 | 0.323 | 0.427 | 0.423 | 0.517 | 0.601 |
| Firmicutes | **0.002** | 0.490 | 0.612 | 0.730 | 0.302 | 0.382 | 0.480 |
| Fusobacteria | **0.002** | 0.000 | 0.000 | 0.000 | 0.000 | 0.003 | 0.016 |
| Proteobacteria | 0.051 | 0.006 | 0.016 | 0.048 | 0.028 | 0.058 | 0.115 |
| Verrucomicrobia | 0.236 | 0.000 | 0.000 | 0.005 | 0.000 | 0.000 | 0.000 |